\documentclass[a4paper,fleqn]{cas-sc}

\usepackage{lipsum}
\usepackage{paralist}
\usepackage{subfigure}
\usepackage{hyperref}
\usepackage[numbers]{natbib}

\begin{document}
\let\WriteBookmarks\relax
\def\floatpagepagefraction{1}
\def\textpagefraction{.001}
\shorttitle{Study of Emergency Evacuation Tactical Decision Making}
\shortauthors{Laura M. Harris et~al.}
\title [mode = title]{Use of immersive virtual reality-based experiments to study tactical decision-making during emergency evacuation}                      
                    


    
\author[1]{Laura M. Harris}[type=editor,
                        role=Graduate Researcher,
                        orcid=0000-0001-8785-8007]

\ead{cty863@vols.utk.edu}


\author[1]{Subhadeep Chakraborty}[type=editor,
   role=PI,
   orcid=0000-0001-5035-9925]
 
\ead{schakrab@utk.edu}

\author[2]{Aravinda Ramakrishnan Srinivasan}[type=editor,
            role=Collaborator,
            orcid=0000-0001-9280-7837]
\cormark[1]  

\ead{A.R.Srinivasan@leeds.ac.uk}

\address[1]{Mechanical Aerospace and Biomedical Engineering, The University of Tennessee, Knoxville}
\address[2]{Institute for Transport Studies, University of Leeds}
\cortext[cor1]{Corresponding author}

\begin{abstract}
Humans make their evacuation decisions first at strategic/tactical levels, deciding their exit and route choice and then at operational level, navigating to a way-point, avoiding collisions. What influences an individuals at tactical level is of importance, for modelers to design a high fidelity simulation or for safety engineers to create efficient designs/codes. Does an unlit exit sign dissuades individual(s) to avoid a particular exit/route and vice versa? What effect does the crowd's choices have on individual's decision making? To answer these questions, we studied the effect of exit signage (unlit/lit), different proportions of crowd movement towards the exits, and the combined (reinforcing/conflicting) effect of the sign and the crowd treatment on reaction times and exit choices of participants in an immersive virtual reality (VR) evacuation experiment. We found that there is tolerance for queuing when different sources of information, exit signage and crowd movement reinforced one another. The effect of unlit exit signage on dissuading individuals from using a particular exit/route was significant. The virtual crowd was ineffective at encouraging utilization of a particular exit/route but had a slight repulsive effect. Additionally, we found some similarities between previous studies based on screen-based evacuation experiments and our VR-based experiment. 
\end{abstract}





\maketitle

\section{Introduction}
The behavior of individuals and crowds is an important topic of research which has helped in preventing crowd disasters and in improving pedestrian flow and safety~\cite{moussaid2018virtual,haghani2018crowd}. Crowd behavior has been studied from different viewpoints such as animal swarms, judgment formation, and consumer behaviors~\cite{helbing2005self,moussaid2009experimental,camazine2003self,castellano2009statistical,moussaid2015amplification}. Individual's decision-making related to emergency evacuation situations has been studied with screen-based experiments in order to understand different factors influencing individuals during evacuation like scenarios~\cite{bode2013human,bode2014human,bode2015information}. Researchers have broadly categorized the decision-making mechanism into strategic, tactical and operational level~\cite{haghani2016pedestrian,hoogendoorn2004pedestrian,haghani2018crowd}. The strategic level is related to when the individual(s) make the decision to start their evacuation, the tactical decisions refers to individual's exit and route choice, and operational level takes care of local interactions like collision avoidance as defined by Warren et al.~\cite{warren2018collective}. In reality, the tactical decisions of individuals have a profound effect on the overall evacuation process for the entire crowd. For example, if every evacuee from a building with multiple exit choices/routes chose to exit via the shortest path in order to optimize chance of successful egress, it can lead to dangerous overcrowding at that particular exit/route and thus eventually become a non-optimal choice for both the individual and the crowd~\cite{10.1007/978-3-319-93372-6_19}.

\subsection{Social Influence}
A crowd's behavior evolves over time and the behavioral propagation can occur through interactions between the individuals within the crowd~\cite{moussaid2017reach} which can lead to interesting crowd dynamics such as lane formation and propensity to choose a certain exit~\cite{helbing2005self}. The social influence of the crowd on individuals can be attractive, neutral, or repulsive as discussed by Warren et al.~\cite{warren2018collective}. The attractive social influence is more generally known as social imitation (or herding)~\cite{helbing2000simulating, haghani2019panic}. It can sometimes lead to sub-optimal performances due to overcrowding of routes and exits~\cite{shiwakoti2017likely}. The overcrowding can lead to increased congestion and exit times~\cite{helbing2000simulating,thompson2018review}. Conversely, individuals have been found to display a repulsive social influence to crowd thus avoiding them. Some studies have found repulsive (avoiding) or neutral influence of the crowd on individuals~\cite{bode2013human,haghani2017social,haghani2019herding,lovreglio2014role} as well. Additionally, queue lengths along with quickest and shortest path to safety have been found to be an influencing factor on route and exit choice~\cite{bode2015information}. 

Lin et al.~\cite{lin2020people} studied effect of social influence in an emergency evacuation situation. They utilized a virtual train station with simulated fire emergency. The study concluded that in an unevenly split $(80-20)$ crowd, individuals tend to follow the bigger crowd. Haghani et al.~\cite{haghani2017stated} compared the stated and revealed choices collected through survey in a virtual evacuation. They found that participants exhibited similar decision pattern in both type of data collection. They also found that people chose the most crowded exit more often. Nilsson et al.~\cite{nilsson2009social} found that social influence is based on distance and plays an important role in the initial evacuation response. They concluded that physically closer people had more influence than people at distance. Additionally, they found that social influence is more important when the evacuation cue is unclear or uninformative. Kinateder et al.~\cite{kinateder2014social} verified that a virtual crowd exerted social influence on participants in a cave immersive virtual reality (VR) system. The virtual crowd also affected route choice of participants. Moussaid et al.~\cite{moussaid2016crowd} found that social imitation was more due to a density effect rather than social imitation. Additionally, they concluded that VR experiment elicited similar decision pattern to real-life~\cite{moussaid2016crowd}. Kinateder et al. found that participants in their VR experiment chose to exit through familiar doors and found that this effect was increased when their virtual neighbors also left through the familiar door~\cite{kinateder2018exit}. Surrounding persons were found to influence evacuee exit decision~\cite{zhu2020follow}. Additionally, Zhu et al. found that strangers affected people similarly to non-strangers~\cite{zhu2020follow}. A stated choice survey was conducted to elicit the effect of social influence and distance to exits by Lovreglio et al.~\cite{lovreglio2014discrete}. Subsequently, the fitted a discrete choice models to better understand exit choices of individuals. Thus, it is clear that crowd has an effect on individual's tactical decision during an emergency evacuation. 

\subsection{Sign Influence}

Static sources such as signs, perceptual access, architectural differentiation, and plan configurations are also important factors in understanding the tactical decision making of individuals during emergency. Building signage are common static directional information source. Exit signs also provide information about emergency exits from a building. Cognition of signs as well as the effectiveness of the content of signs were studied and applied to exit designs~\cite{collins1979evaluation, lerner1981evaluation, collins1982development, collins1983evaluation}. The size of the content was studied and larger content was found to be more effective~\cite{boyce1995effective}. The color of the content for the signs were studied and green or red lettering have been found to increase recognition distance~\cite{jouellette1988exit}. Sign design and illumination have been found to be important for sign visibility~\cite{jin2002visibility}. Signs which update have been found to be more trusted than fully static signs~\cite{meier2015influence}. Comprehensive knowledge of the environment was found to be not necessary for reasonable decisions to be made~\cite{passini1984spatial}.

Fu et al.~\cite{fu2019influence} studied how signs affect people with and without the presence of a crowd. Signs were found to be more effective in the absence of crowd influence. Additionally, the average decision time of those who did not follow signage was not atypically out of normal. Olander et al.~\cite{olander2017dissuasive} found that dissuasive signs, like a red X added on the sign to indicate no entry, to be effective in providing directional information. Tang et al.~\cite{tang2009using} summarized that exit signage were important for way-finding decisions, but individuals do not always follow them. Importantly, evacuations were slower when no signage were present~\cite{tang2009using}. Bode et al.~\cite{bode2014human} utilized an interactive $2D$ VR experiment and found that signs have a significant influence on exit choice of participants. Kinateder et al.~\cite{kinateder2019color} found that green colored signs were most attractive for exit utilization. The influence of visual information by means of exit signs and corridor illumination were studied by Dachner et al. with a small pool of participants and both were found to influence emergency decision-making~\cite{dachner2016effects}. Galea et al.~\cite{galea2014experimental} found that dynamic exit signage increased the visibility of the exit signs from $38\%$ to $77\%$ compared to static exit signage. Galea et al.~\cite{galea2017evaluating} found that dissuasive signage with reinforcement from a voice alarm system successfully redirected people towards the optimal exit by $66\%$.

\subsection{Use of VR in Egress Literature}

Researchers have used methods such as surveys~\cite{shiwakoti2017likely,lovreglio2014role} and evacuation drills~\cite{shields2000study,huo2014investigation} to study interactive evacuation scenarios~\cite{haghani2017social,bode2014human,zhu2016experimental, fu2019influence, heliovaara2012pedestrian}. More recently, immersive VR has been a promising avenue of information gathering which minimizes risks and cost associated with evacuation experiments. VR has been used in many fields to study human factors and has been found to be an effective tool to study real world behaviors~\cite{kobes2010exit}. Several studies have found that VR elicits similar non-emotional responses as real-world experiments~\cite{peperkorn2014triggers, malthe2012virtual, johansson2013utrymning, tornros1998driving, hirata2007development, shechtman2009comparison} and adequate emotional responses as situation demands~\cite{kinateder2014social, calvi2010analysis, calvi2011long, muhlberger2007virtual}.

There are also known limitations in ergonomics and technology which can prevent complete reliability of the VR experiment e.g. unrealistic AI, unrealistic movement, unrealistic environments, unrealistic hazards, motion sickness~\cite{feng2018immersive}, and lack of inherent danger. But, some validation has been performed on VR experiment by comparing results from VR to real-world data. Kobes et al.~\cite{kobes2010exit} performed a validation study for use of serious games for behavioral evacuation studies where VR-generated data was compared to real-world data and found to elicit similar behavior. Kinateder et al.~\cite{kinateder2016social} found some interesting correlation between real world and VR experiments. Reactions to an alarm were reduced, a comparable response to positive influence from bystanders, and a weaker response to negative influences where observed in VR. Kinateder et al.~\cite{kinateder2018crowd} provided an in-depth comparison of real world and virtual world behaviors and the capabilities of VR.

In summary, it is important for modelers and safety engineers to account for all the different factors that can influence an individual's tactical decision. What factors can encourage individuals to take a particular route, what factors discourage individuals to take a particular route, does the crowd movement influence the decision making, if yes, whether it is a positive influence or a negative influence, does the exit signage being lit or unlit have opposite effect on the individual's choice? Also, does virtual reality (VR) based experiment elicit different response compared to screen-based experiments? Bode et al.~\cite{bode2013human,bode2014human,bode2015information} have performed comprehensive analysis of factors affecting the tactical decision making with a screen-based virtual evacuation experiment. We are interested in comparing the results from them with data collected in an immersive, first-person view, VR-based experiment. More particularly, the current study expands on the work reported by Bode et al.~\cite{bode2014human} on directional information source on exit choice of humans with 2D virtual experiment. Specifically, this study attempts to address how lit and unlit exit signs and crowd configuration affect evacuee exit choice behaviors in an immersive VR emergency evacuation simulation. The primary differences between this work and that of Bode et al. were the use of immersive VR environment, the addition of an uneven crowd configuration and the lit and unlit exit signage. 
 
\begin{figure*}[b]
 \centering
 \includegraphics[scale=0.35]{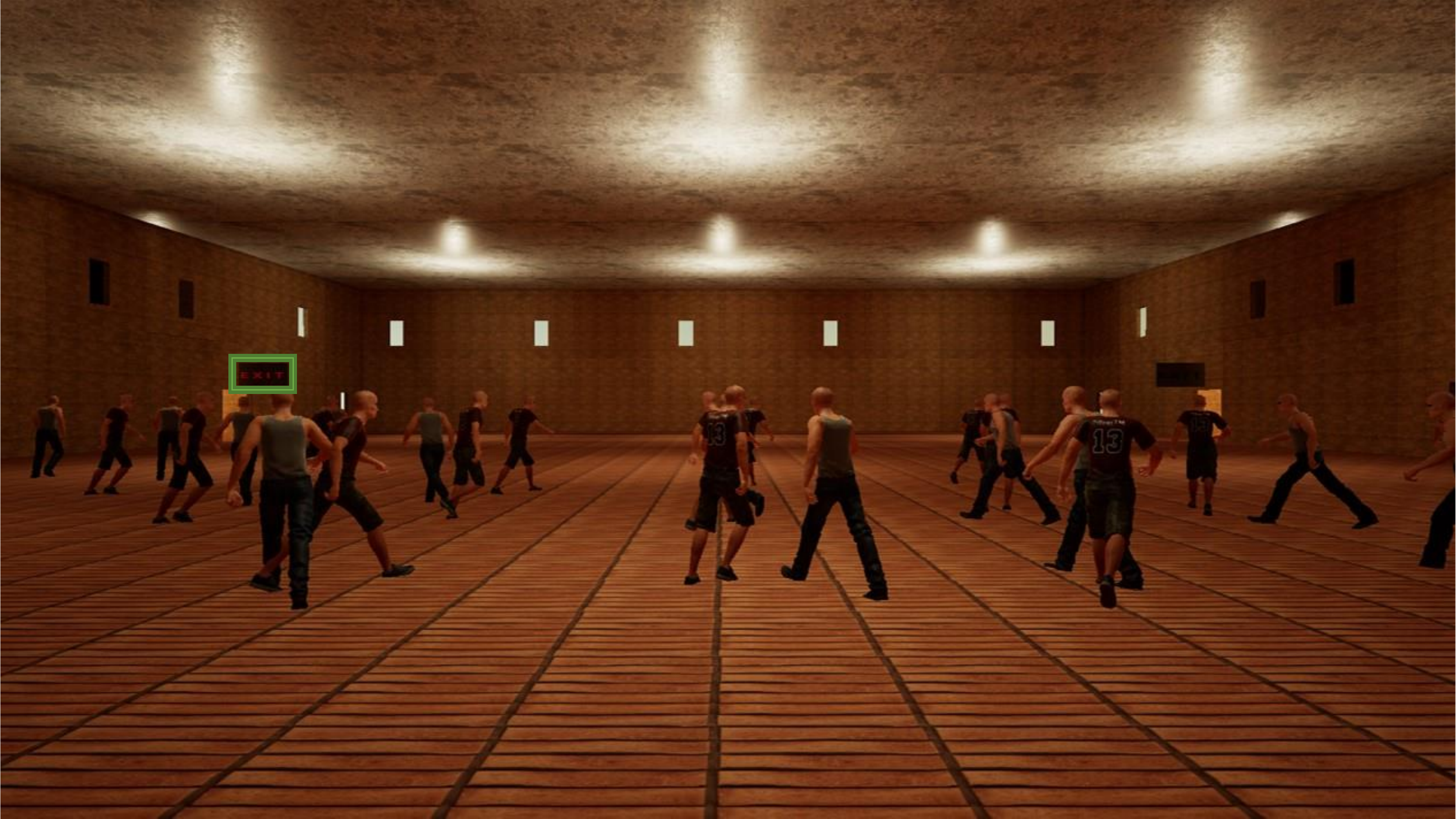}
 \caption{A screen capture of the immersive environment: Participant perspective of scenario 2: Sign}
 \label{fig:Egress2}
\end{figure*}
\section{Methods}
\begin{figure*}[t]
 \centering
 \subfigure[]{
  \includegraphics[scale=0.5]{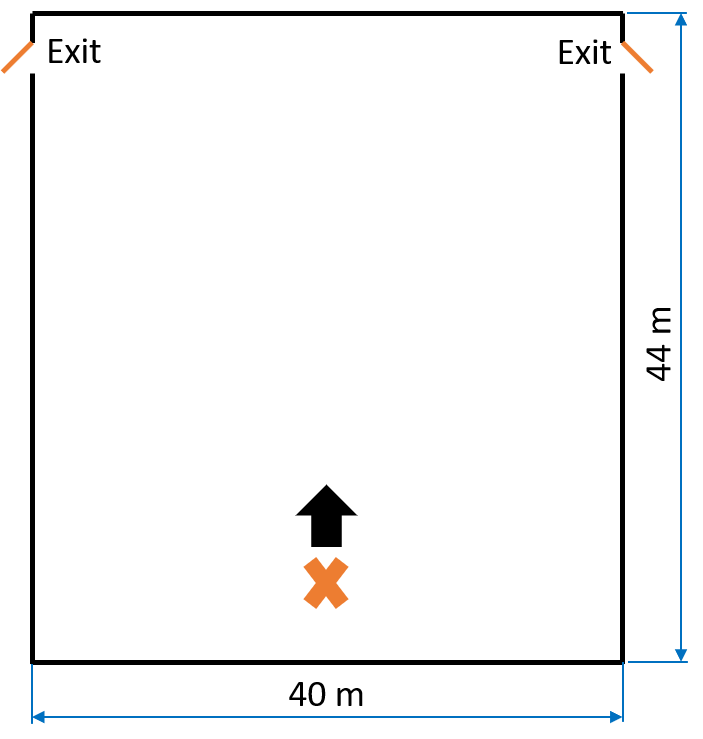}
 \label{fig:Room1}}
 \subfigure[]{
 \includegraphics[scale=0.5]{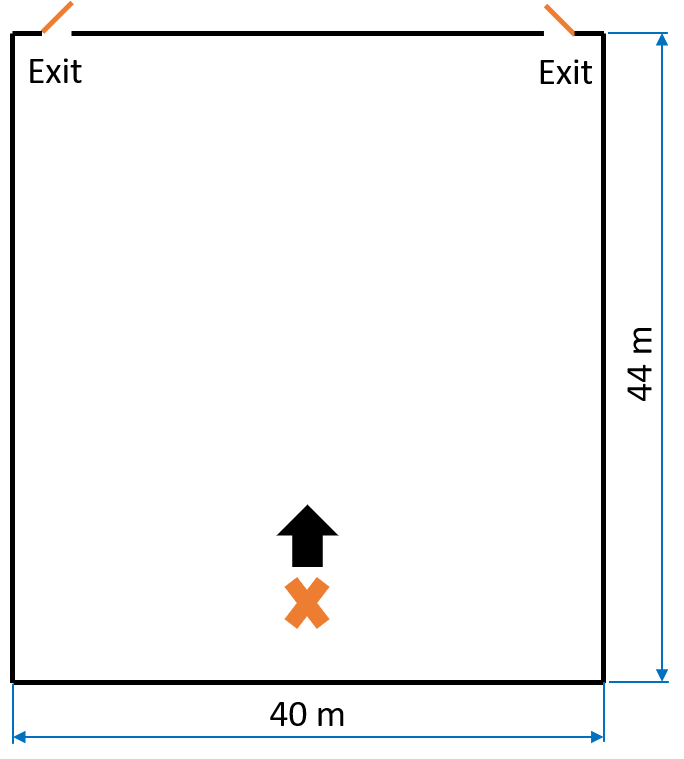}
 \label{fig:Room2}}
 
 \caption{2D Representation of Experimental Rooms. All participants had their VR avatar start at the equidistant position from the exit as marked by the orange X facing towards the exits as indicated by the arrow mark in all scenarios}
 \label{fig:2DRooms}
\end{figure*}
\subsection{Experimental Design}\label{Sec:Expt_Design}
Unreal Engine 4 was used to develop the immersive VR environments. A head-mounted display, a HTC Vive Pro (with 1440x1600 pixels resolution per eye), was used to provide the participants with a fully immersive experience. Participants navigated their virtual avatar in the simulated environment with a joystick. A simple room layout was preferred to explore the effect of directional information provided by crowd and exit signage on exit choice. This is similar to the simple room layout utilized in literature~\cite{bode2013human,bode2014human,bode2015information}. The room allowed the participants full visual access to both exits and exit signage. An example room (\figurename~\ref{fig:Egress2}) and a 2D representation of two representative room layouts is provided in \figurename~\ref{fig:2DRooms}. The room was $40$ meters wide and $44$ meters long. The exit signs were $1$ meter wide and $0.4$ meters in height. The signage were designed to be bigger than the standard dimensions to counter the loss of clarity due to pixelation in the HTC Vive screen. We prioritized being able to read the signs over strictly adhering to sizing standards. The exit doorways were 2 meters wide which conforms with the US Department of Labor's Occupational Safety and Health Administration (OSHA) guidelines for Emergency Exit Routes (osha.gov)~\cite{OSHAFact}. The exits were located either against the far wall from the participant starting point or at the end of the side walls. The participants' start location was in the central bottom portion of the room (marked with an ``X'') as can be seen in 2D representation of the room in \figurename~\ref{fig:2DRooms}. Furthermore, a crowd of co-evacuees consisting of programmed non-player characters (NPCs) was positioned around the participant within the participants' virtual field-of-view rather than throughout the room. A total of six scenarios were designed. The scenarios are summarized in \tablename~\ref{tab:scenParam}. All scenarios had two exits (the left and the right exit) and each participant started from a position equidistant from the exits as indicated in \figurename~\ref{fig:2DRooms}.

\begin{table*}[b]
\centering
	\caption{Experimental parameters for each of the 6 scenarios tested.}
	\label{tab:scenParam}
	\begin{tabular}{|lllll|}
		\hline
		\multicolumn{1}{|l|}{\textbf{\#} } & \textbf{Scenario} & \textbf{Lit Signs} & \textbf{Crowd: Left} & \textbf{Crowd: Right}\\ \hline
		\multicolumn{1}{|l|}{1} & Control & Neither & 25 & 25 \\
		\multicolumn{1}{|l|}{2} & Sign & Left Only & 25 & 25 \\
		\multicolumn{1}{|l|}{3} & Crowd & Neither & 0 & 50 \\
		\multicolumn{1}{|l|}{4} & Sign+Crowd & Left Only & 50 & 0 \\
		\multicolumn{1}{|l|}{5} & Sign-Crowd & Right Only & 50 & 0 \\
		\multicolumn{1}{|l|}{6} & Uneven Crowd & Neither & 15 & 35 \\
 \hline
	\end{tabular}
\end{table*}

The appearances of the NPCs in VR experiments have been found to not have any significant effect on decision-making by Llobera et al.~\cite{llobera2010proxemics} and Bruneau et al.~\cite{bruneau2015going}. Thus, all NPCs in our experiments appeared as males and were identical to each other. The NPCs used simple way-finding to find the closest path to their designated exit. The NPCs and participant moved at the same speed of 1.5 meters per second (3.4 miles per hour) for a faster than normal walking speed for the evacuation~\cite{buchmueller2006parameters}. \tablename~\ref{tab:scenParam} details the experimental parameters for each of the 6 Scenarios tested. Scenario 1 was designed to be the control, scenario 2 was the base sign control test, scenario 3 was designed to study the effect of the crowd on the decision-making, scenario 4 was a reinforcing-treatments scenario with crowd and sign in agreement, scenario 5 was the conflicting-treatments scenario and scenario 6 was an uneven crowd treatment scenario with an unevenly split crowd. 

The NPCs acted as a potential influence on the participants' exit decision. To further elicit the crowding behaviors during an evacuation, the NPCs were programmed to not have collision avoidance with other evacuees to allow crowding at exits. It is to noted that the NPCs can never overlap but can be blocked by each other when we mean they can collide. The NPCs were randomly assigned to an exit according to the scenario description. Additionally, the exits were visible in all scenarios. The participants could not see into the room beyond the exits. One or both the exit signs were unlit to provide a dissuasive sign effect to the participants according to the scenarios. The dissuasive effect is hypothesized as lit exit signs are expected whereas unlit exit sign(s) were present thus biasing the participants not to use that particular exit. We refer to this as an updated sign in this work. A klaxon alarm activated in the virtual environment when the evacuation scenario started and ended when the participants' avatars exited the room.

The order that the participants performed the scenarios was randomly chosen and tracked. A within subjects design was used. The number of participants who saw a given scenario as their first scenario, as opposed to later in the experiment, are provided in \tablename~\ref{tab:FirstNotFirst}. The participants had the opportunity to complete all six scenarios. Participants who performed multiple scenarios may have had a source of self-feedback~\cite{mather2012risk} from prior scenarios, but this was found to not be a significant influence in~\cite{lin2020people}. The path and the exit choice of the participants were recorded. In order to avoid participants expecting the same scenarios, the room for each scenario was given unique wall, floor and ceiling textures. Additionally, two sets of room layouts were used in an attempt to further reduce habituation as seen in \figurename~\ref{fig:2DRooms}. 
\begin{table*}[t]
\centering
	\caption{Summary of scenarios which were seen as the first scenario and those which were not seen as the first.}
	\label{tab:FirstNotFirst}
	\begin{tabular}{|lllll|}
		\hline
		\multicolumn{1}{|l|}{\textbf{\#} }& \textbf{Scenario} & \textbf{First} & \textbf{Not First} & \textbf{Total}\\ \hline
		\multicolumn{1}{|l|}{1} & Control & 11 & 46 & 57 \\
		\multicolumn{1}{|l|}{2} & Sign & 10 & 45 & 55 \\
		\multicolumn{1}{|l|}{3} & Crowd & 10 & 44 & 54 \\
		\multicolumn{1}{|l|}{4} & Sign+Crowd & 10 & 44 & 54 \\
		\multicolumn{1}{|l|}{5} & Sign-Crowd & 10 & 43 & 53 \\
		\multicolumn{1}{|l|}{6} & Uneven Crowd & 10 & 47 & 57 \\
 \hline
	\end{tabular}
\end{table*}

\subsection{Data Analysis Approach}

The three analyzed summary statistics were 1) the participants' probability to follow directional information provided by the environment in the form of signage and crowds' exit choice, 2) the participant's desire to change their exit choice after initial decision, and 3) the time they took to make their initial decision to choose an exit.

To determine the probability that the participants changed their exit decision, P(change decision), the event "change decision" was defined as where the participant started to walk towards an exit and then began to walk to the other exit after moving at least one-fifth the lateral distance in the direction of the former exit. This is consistent with the definition by Bode et al.~\cite{bode2014human} in literature. The probability that the participant went to the exit with a particular treatment, P(follow treatment), was found from the proportion of the population which chose the exit indicated by the treatment. When two treatments reinforced each other, e.g. (Sign+Crowd), the summary statistics for each treatment was equal to the probability to follow the treatment, P(Follow Sign)=P(Follow Crowd)=P(Follow Treatment). When two treatments conflicted with each other, e.g. (Sign-Crowd), the follow treatment summary statistic for each treatment was found by the probability of the participants to choose the exit of the respective treatment. The decision time was chosen as the time from the start of the simulation until the instant when the participant started to move towards an exit. Since the start position of the participant's avatar was set by the simulation, the start of the movement could be identified by tracking the participant's avatar position in post-processing. The initial exploratory movements such as turning to search for exits or looking at the NPCs were thus not considered to be an actual determined movement towards exit, but rather counted as part of the decision time.

The significance level of $95\%$ was chosen for all models and tests performed. Cochran's Q-tests \footnote{Ben Jann, 2004.
"COCHRAN: Stata module to test for equality of proportions in matched samples (Cochran's Q)," Statistical Software Components
S4444105, Boston College Department of Economics, revised 27 Oct 2004.
\href{https://ideas.repec.org/c/boc/bocode/s444105.html}{https://ideas.repec.org/c/boc/bocode/s444105.html}} were used to determine if the follow treatment and change decision statistics had significantly different proportions for each relevant scenario. A pairwise comparison between the scenarios and Control, Sign, and Crowd were performed to test desired comparisons ex ante. A Sidak correction was used. A one-way repeated measures ANOVA was performed to determine if there was significance across participants and across scenarios for decision time. After performing the ANOVA, a pairwise comparison between scenarios was performed with a Sidak correction. 

The data was also split into the participants who performed a given scenario first and those who had completed the same scenario after completing another scenario. The data was analyzed for each scenario separately to discover learning effects from observation of the means and standard deviations. 

\subsection{Participants}

\begin{table*}[t]
	\caption{Demographics Summary}
	\label{tab:demoSummary}
	\centering
	\begin{tabular}{|lll|}
		\hline
		\textbf{Age Demographics} & & \\ \hline
		\multicolumn{1}{|l|}{Age (years)} & Frequency & Percent (\%) \\ \hline
		\multicolumn{1}{|l|}{18-24} & 27 & 44.26 \\
		\multicolumn{1}{|l|}{25-34} & 17 & 27.87 \\
		\multicolumn{1}{|l|}{35 and over} & 16 & 26.23 \\
		\multicolumn{1}{|l|}{NA} & 1 & 1.64 \\
		\multicolumn{1}{|l|}{Total} & 61 & 100 \\ \hline
		\textbf{Gender Demographics} & & \\ \hline
		\multicolumn{1}{|l|}{Gender} & Frequency & Percent (\%) \\ \hline
		\multicolumn{1}{|l|}{Man} & 42 & 68.85 \\
		\multicolumn{1}{|l|}{Woman} & 18 & 29.51 \\
		\multicolumn{1}{|l|}{Other} & 1 & 1.64 \\
		\multicolumn{1}{|l|}{Total} & 61 & 100 \\ \hline		
	\end{tabular}
\end{table*}

Prior to any data collection, approval was sought from the local ethics committee, the Institutional Review Board (IRB), University of Tennessee, Knoxville. This study was approved by the IRB (approval number UTK IRB-17-04159-XM). The data collection was conducted prior to the Covid-19 pandemic. The data collection itself was anonymized. No personally identifiable information were recorded. The data have been stored according to local IRB recommendations. 

Participants were recruited from the staff and students at the University of Tennessee, Knoxville campus. No incentives were provided for participation. A total of 64 participants were recruited through an open call for volunteers. However, only the data of 61 participants was used. Participants were eligible if they were 18 years or older and had no known sickness to VR-based experiments. The demographics of the participants are listed in \tablename~\ref{tab:demoSummary}. For analysis purpose, gender was converted to 0 for participants who listed themselves as man and 1 for participants who listed themselves as woman.

At least $10$ participants were desired to perform each scenario as the first one and at least $30$ participants total for each scenario given the precedence set by previous works~\cite{dachner2016effects, johansson2013utrymning}.

\subsection{Procedure}\label{Sec:Procedure}
Each participant was provided with an information sheet about the study and a consent form, both approved for this specific experiment by the local ethics committee. The participants had opportunity to discuss their participation and potential benefit (to the scientific community) with the researchers before they signed the consent form. After signing the informed consent form, each participant was given a demographic survey. Next, the participants were asked to wear an HTC Vive Pro virtual reality headset and placed in a "VR preparatory setup" which consisted of 2 rooms surrounded by corridors and several exits. The participant started from one of the rooms with an exit and an exit sign above it. Upon leaving this room, the participant made a series of left or right decisions to exit out of the building. The participants then repeated this same scenario except with the addition of NPCs who also were egressing the building. No data from the preparatory scenarios was used in our analysis in this work, but it provided an opportunity for participants to get comfortable with the immersive VR environment and the navigation control with the joystick.

After the VR preparatory environment, the participants then completed as many scenarios in the actual experimental setup (described in subsection~\ref{Sec:Expt_Design}) as they were comfortable with. The number of scenarios varied between participants due to their comfort with the VR and time involved. The 1-room simplified environment, reported in this paper, was designed to test fundamental exit choices rather than more complex time-dependent tests for memory, familiarity, etc. This was done in an attempt to limit the VR exposure time, which can cause motion sickness, but still provide the time needed to develop the feeling of involvement and engagement in the evacuation process. Moreover, a warning siren sound played through the VR headphones throughout the experiment, to provide a sense of urgency during the experiment.

Finally, when the participants completed as many scenarios as possible, they were asked to complete an exit survey. The participants were asked to rate how different factor affected their decision-making during a scenario. The rated factors were 1) the exit sign lit, 2) perceived time-to-exit, 3) follow crowd, 4) avoid crowd, and 5) previous exit choice(s). Lit sign was provided for participants in the rating to inform us about how effectiveness of exit sign lighting versus non-lighting in influencing their exit choice. Time-to-exit was provided for participants to describe if the time to leave the room was an important factor in determining their exit decisions. Both follow crowd and avoid crowd were provided to ascertain the effect of the crowd on individuals. The previous exit choice(s) was provided to determine what effect the previous exits decision (when participating in multiple scenarios) had on the participant's exit choice. The rating scale was from 1 (lowest) to 5 (highest). Additionally, they were requested to elaborate their thought process behind the ratings. A screenshot of the exit survey is provided in \figurename~\ref{fig:eSurvey}. The experimenter's script for the entire data collection process is provided in Appendix~\ref{sec:Appx1}. 
\begin{figure*}[t]
 \centering
 \includegraphics[scale=0.4]{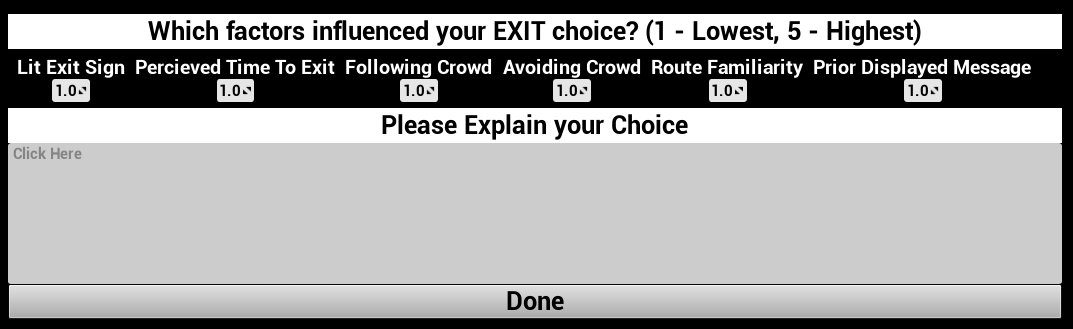}
 \caption{A screenshot of participant's exit survey}
 \label{fig:eSurvey}
\end{figure*}


\section{Results}

\subsection{Survey}
\begin{figure*}[t]
    \centering
    \includegraphics[width = 0.5\linewidth]{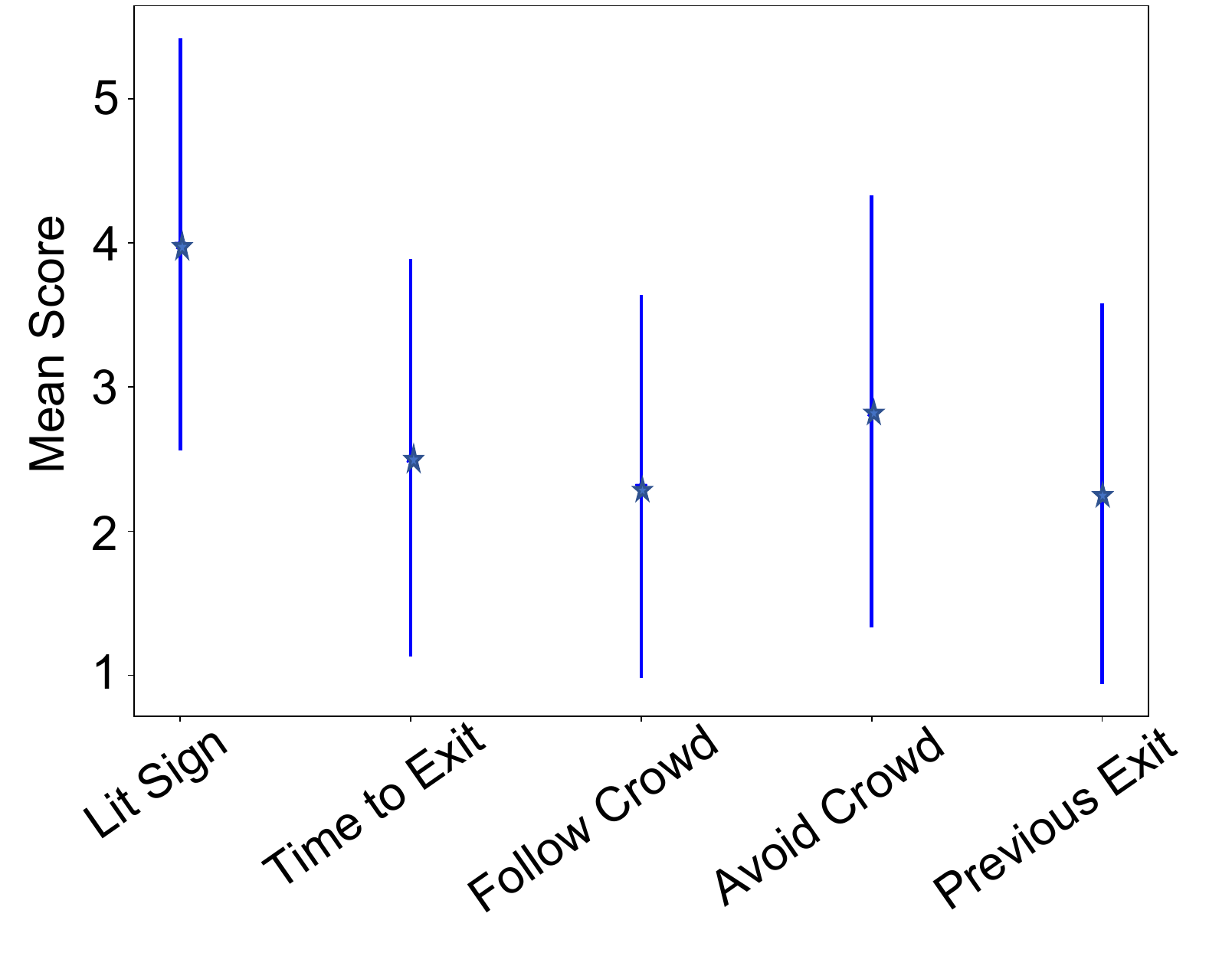}
    \caption{A plot visualizing the mean score for each of the influencing factors along with their standard deviation. }
    \label{fig:Survey}
\end{figure*}

Fifty of the sixty-four participants completed the survey. The studied factors were lit sign, time-to-exit, follow crowd, avoid crowd, and previous exit choice(s). The participants were asked to specify how much they thought each factor affected their exit choices. The values ranged from 1 as the lowest and 5 as the highest possible scores. The averages for the survey responses are provided in \figurename~\ref{fig:Survey}. The lit sign was listed as the most influential followed by avoid crowd, time-to-exit, follow crowd, and previous exit choice(s) as the least influential. 
\begin{table*}[b]
\centering
	\caption{Chose Left Summary Statistics: First vs. Not First}
	\label{tab:CLSumStatFvNF}
	\begin{tabular}{|llll|l|}
		\hline
		\multicolumn{1}{|l|}{\textbf{\#}} & \textbf{Scenario} & \textbf{First (\%)} & \textbf{Not-First (\%)} & \textbf{Combined (\%)}\\ \hline
		\multicolumn{1}{|l|}{1} & Control & $27.3$ & $54.3$ & $49.1$\\
		\multicolumn{1}{|l|}{2} & Sign & $80.0$ & $82.2$ & $81.8$\\
		\multicolumn{1}{|l|}{3} & Crowd & $20.0$ & $54.5$ & $47.3$\\
		\multicolumn{1}{|l|}{4} & Sign+Crowd & $100.$ & $72.7$ & $77.8$\\
		\multicolumn{1}{|l|}{5} & Sign-Crowd (Sign) & $40.0$ & $16.3$ & $21.8$\\
		\multicolumn{1}{|l|}{6} & Uneven Crowd & $40.0$ & $46.8$ & $43.9$\\
 \hline
	\end{tabular}
\end{table*}

\subsection{First vs. Not-First Participant Results}
The first versus not-first data for the chose left statistics were provided in \tablename~\ref{tab:CLSumStatFvNF}. The chose left statistics refers to the choice to take the left exit from the participant's perspective. The primary difference in the first versus not-first results was observed in Control and Crowd scenarios. Small differences occurred in the other scenarios but the behaviors of the participants is similar for the first and not-first scenarios. The probability of the participants to choose an exit was not significantly different from chance when the not-first and first scenario participant data was combined for the Control scenario; however, the participants who saw Control as their first scenario chose the left exit $27.2 \%$ of the time and those that did not see the control as their first scenario chose the left exit $54.3 \%$ of the time. Eleven participants saw Control as their first scenario compared to forty-six who saw it after their first scenario. The participants who saw Crowd as their first scenario chose to avoid the crowd (left) $20.0 \%$ of the time, but those that saw Crowd after their first scenario chose to avoid the crowd $54.5 \%$ of the time. Ten participants saw Crowd as their first scenario and forty-four participants saw it after their first scenario. These differences indicate a potential right bias in the case of the Control and a repulsive effect of the crowd when the participants had been exposed to previous scenarios. 
\begin{figure*}[t]
 \centering
 \includegraphics[width = 0.5\linewidth]{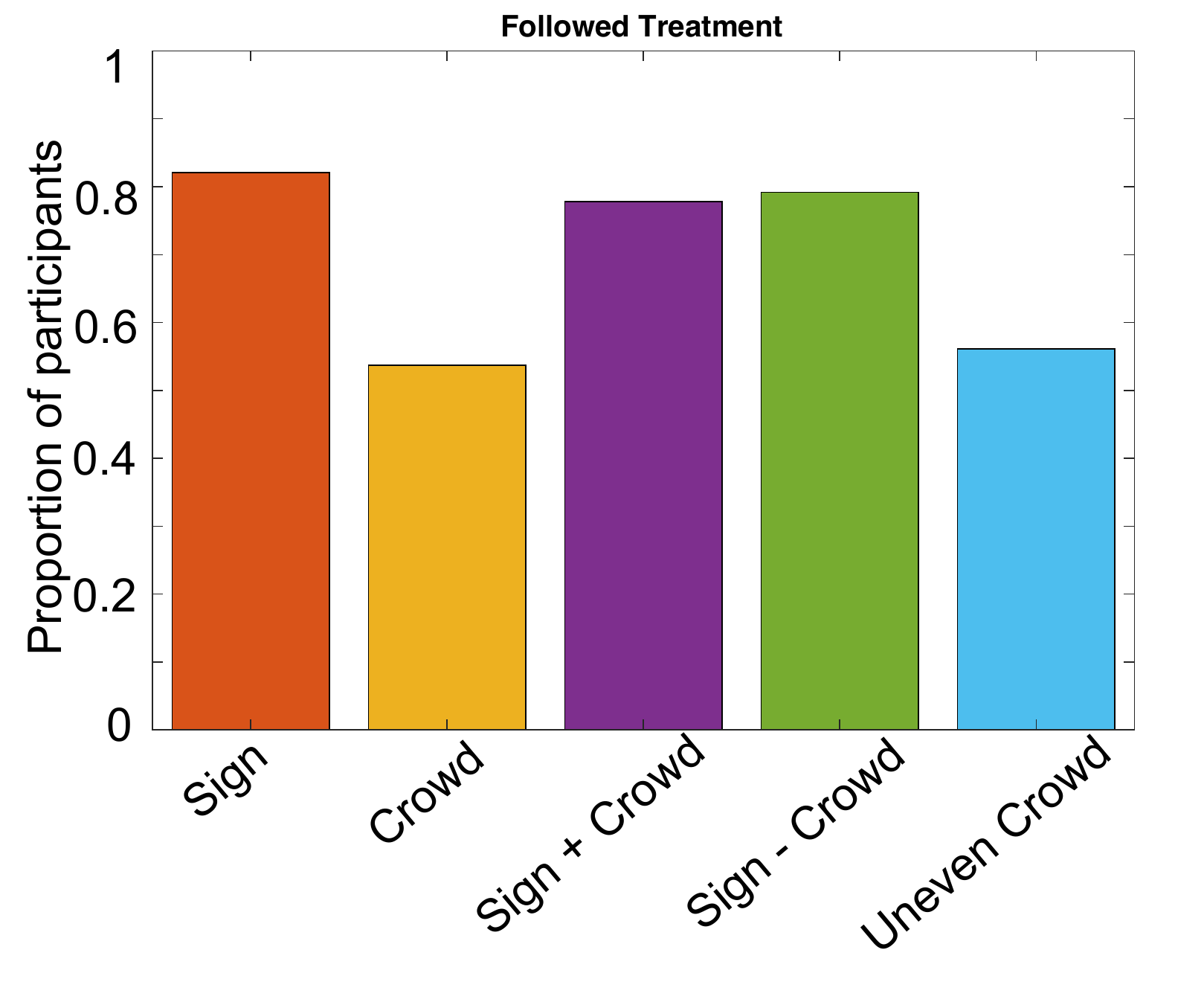}
 \caption{Proportion of participants who followed the respective treatment(s) (combined percentage)}
 \label{fig:FollowStat}
\end{figure*}
 \begin{figure*}[b]
 \centering
 \includegraphics[scale=0.6]{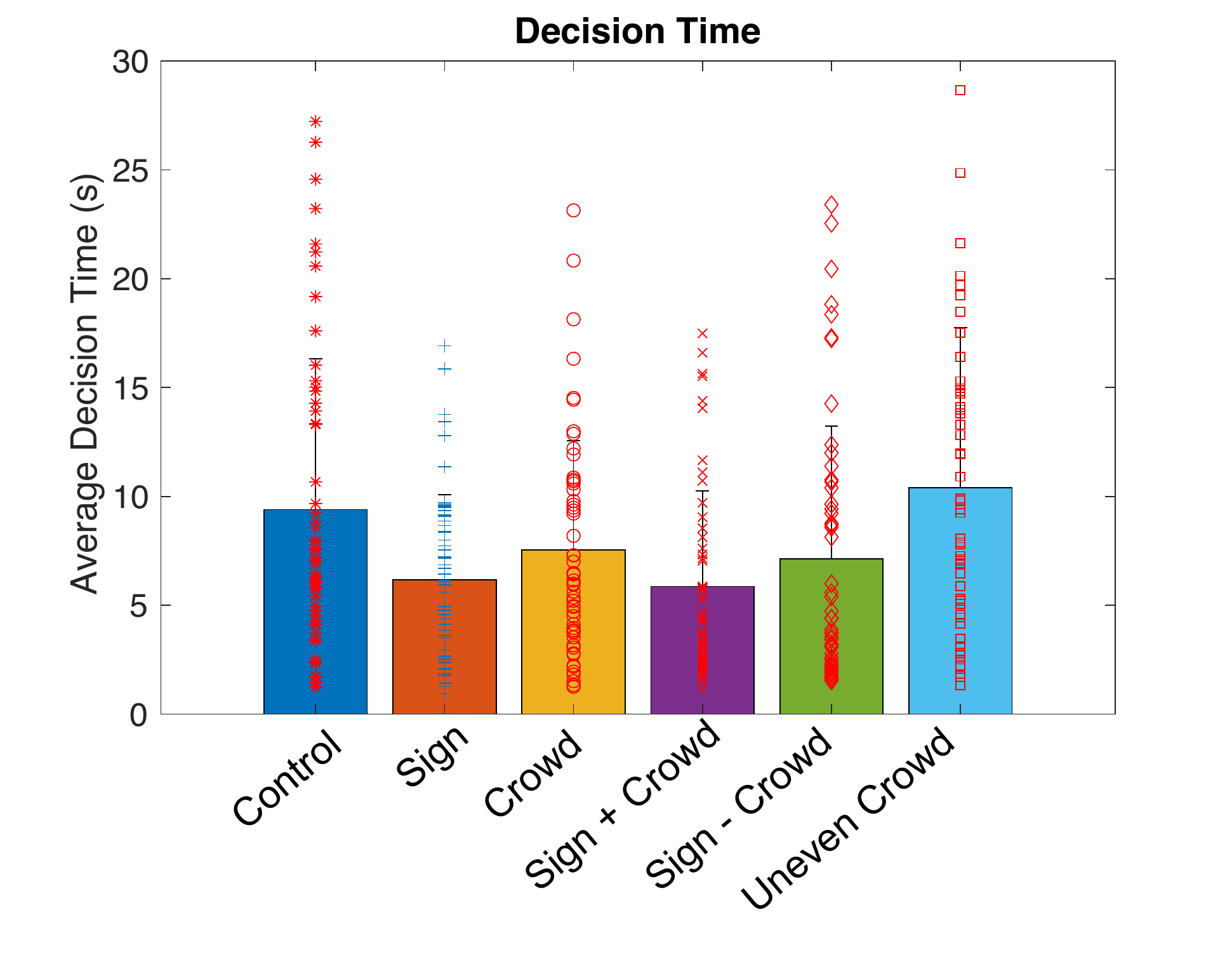}
 \caption{Average decision time for each scenario}
 \label{fig:DecisionTime}
\end{figure*}
\begin{table*}[t]
\centering
	\caption{ Decision Time Summary Statistics: First vs. Not First}
	\label{tab:RTSumStatFvNF}
	\begin{tabular}{|llll|l|}
		\hline
		\multicolumn{1}{|l|}{\textbf{\#}} & \textbf{Scenario} &\textbf{First (s)} &\textbf{Not-First (s)} &\textbf{Combined (s)}\\ \hline
		\multicolumn{1}{|l|}{1} & Control & $5.12\pm2.72$ & $10.4\pm7.24$ & $9.39\pm6.92$\\
		\multicolumn{1}{|l|}{2} & Sign & $5.82\pm2.86$ & $6.23\pm4.13$ & $6.16\pm3.91$\\
		\multicolumn{1}{|l|}{3} & Crowd & $7.48\pm5.19$ & $7.65\pm5.01$ & $7.55\pm5.01$\\
		\multicolumn{1}{|l|}{4} & Sign+Crowd & $6.20\pm2.66$ & $5.66\pm4.67$ & $5.86\pm4.40$\\
		\multicolumn{1}{|l|}{5} & Sign-Crowd & $10.2\pm5.40$ & $6.41\pm6.07$ & $7.13\pm6.09$\\
		\multicolumn{1}{|l|}{6} & Uneven Crowd & $10.1\pm7.82$ & $10.4\pm7.32$ & $10.4\pm7.34$\\
 \hline
	\end{tabular}
\end{table*}

The first versus not-first data for the decision time were provided in \tablename~\ref{tab:RTSumStatFvNF}. Upon visual inspection, the data appeared to be within standard deviation for each scenario as seen in \tablename~\ref{tab:RTSumStatFvNF}. This would indicate that a learning effect likely did not apply to the decision time. The proportion of participants who followed the treatment for each scenario is presented in \figurename~\ref{fig:FollowStat}. The proportion of those who followed the treatment for Sign-Crowd was given to those who followed the sign rather than the crowd. 

\subsection{Decision Time Results}
\begin{table*}[t]
\centering
	\caption{ANOVA Summary}
	\label{tab:Anova}

	\begin{tabular}{|l|llll|}
		\hline
		{\textbf{Variable}} &\textbf{SS} &\textbf{df} &\textbf{MS} &\textbf{F}\\
		\hline
		{Scenario \#} &$4778$ &$5$ &$185.1$ &$8.09$ \\
		& & & &\\
		{Residual} &$6013$ &$263$ &$22.86$ & \\
 \hline
	\end{tabular}
\end{table*}
\begin{table*}[t]
\centering
	\caption{Pairwise Decision Time Comparison Summary}
	\label{tab:PW_DT}
	\begin{tabular}{|l|lll|}
		\hline
		{\textbf{Scenario}} &\textbf{Contrast} &\textbf{Std. Err.} &\textbf{t}\\
		\hline
		{Sign vs Control} &$-3.29$ &$0.914$ &$-3.60$ \\
		{Sign+Crowd vs Control} &$-3.63$ &$0.924$ &$-3.93$ \\
 {Uneven Crowd vs Sign} &$4.28$ &$0.915$ &$4.68$ \\
 {Uneven Crowd vs Crowd} &$3.06$ &$0.921$ &$3.33$ \\
 {Uneven Crowd vs Sign+Crowd} &$4.63$ &$0.926$ &$5.00$ \\
 {Uneven Crowd vs Sign-Crowd} &$3.49$ &$0.918$ &$3.81$ \\
 \hline
	\end{tabular}
\end{table*}
A summary of the decision time can be found in \figurename~\ref{fig:DecisionTime}. A one-way repeated measure ANOVA was used to determine if there were significant differences between the means of the treatment scenarios and across participants for the decision time. The results can be found in \tablename~\ref{tab:Anova}. There was a significant effect of scenario number on decision time, $F(5, 263) = 8.09, p = 0.0000$, so we reject the null hypothesis that the means of decision time are equal between scenarios. Therefore, a pairwise comparison with a Sidak correction was used. No ex ante relationships were assumed between scenarios and decision time. The significant results were between Sign vs Crowd, Sign-Crowd vs Control, Uneven Crowd vs Sign, Uneven Crowd vs Crowd, Uneven Crowd vs Sign+Crowd, and Uneven Crowd vs Sign-Crowd as summarized in \tablename~\ref{tab:PW_DT}. The contrasts found agree with the statistics provided in \tablename~\ref{tab:RTSumStatFvNF}. This seems to indicate that the contrasting treatments produced hesitancy compared to the single or reinforcing treatments. Additionally, decision time is the lowest for scenarios with the sign treatment, (Sign, Sign+Crowd, and Sign-Crowd). 

\begin{table*}[t]
\centering
	\caption{Pairwise Follow Information Comparison Summary}
	\label{tab:PW_FI}

	\begin{tabular}{|l|lll|}
		\hline
		{\textbf{Scenario}} &\textbf{$\chi^2$} &\textbf{p-value} &\textbf{Sidak} \\
		\hline
		{Sign vs Control} &$10.8$ &$0.0014$ &$0.0167$ \\
		{Sign+Crowd vs Control} &$8.33$ &$0.0059$ &$0.0462$ \\
 {Sign-Crowd (Sign) vs Control} &$9.14$ &$0.0037$ &$0.0400$ \\
 {Crowd vs Sign} &$9$ &$0.0041$ &$0.0403$ \\
 {Crowd vs Sign-Crowd (Sign)} &$8.91$ &$0.0043$ &$0.0403$ \\
 \hline
	\end{tabular}
\end{table*}

\subsection{Following Information Results}
A Cochran's Q-Test was used to determine if the null hypothesis that the proportions of following information between scenarios were equal. The test reported that significantly different proportions existed for follow information between the different scenarios, $\chi^2 = 30.3, p = 0.0000$. It was desired ex ante to compare all relevant scenarios with different baseline scenarios. The baselines scenarios were Control, Sign, and Crowd. Therefore, a pairwise comparison with a series of Cochran's Q-Tests with a Sidak correction was used. As summarized in \tablename~\ref{tab:PW_FI}, [Sign vs Control, Sign+Crowd vs Control, Sign-Crowd (Sign) vs Control, Crowd vs Sign, and Crowd vs Sign-Crowd (Sign)] were significantly different. All of the significant pairwise comparisons included the sign treatment.

\section{Discussion}
It is interesting to analyze how each scenario affected the participants' probability to follow the treatment information, to change their decision, and to quickly make their decision. While the Control scenario did not have treatment information, it was expected to have an equal proportion of exits to be chosen. This was not observed as the participants chose the right exit more than the left exit, but only slightly so. The deviation from the expected value was not significant but it may indicate a bias in the participants to choose the right exit. It may be conjectured that this bias is a result of right-handedness of the participants or the layout of the joystick rather than any other factors. This handed bias was found in a work by Veeraswamy et al.~\cite{veeraswamy2011wayfinding} that studied way-finding in buildings. 
 
When the participants saw the Sign-Crowd scenario as their first one, their preference to choose the exit with sign(right exit) or the crowd(left exit) was approximately evenly distributed. However, a drop in following crowd(left exit) was observed when participants saw the Sign-Crowd as a non-first scenario. This is indicative of a learned repulsive effect of the crowd. This may be related to the effect of the physicality of the crowd on the participant while attempting to exit. Furthermore, the survey responses found that the sign was the most important factor on average and avoiding the crowd was the second most important factor. These responses agree with the observed trends in the collected trajectory/exit choice data. Another potential relationship was that the time-to-exit was the third most important factor reported on the surveys, this may indicate that the participant wanted to avoid the crowd to egress faster through the less crowded exit. 

Participants generally chose to follow the lit exit sign over any other treatment, and did so consistently for all scenarios with the lit sign treatment. This agrees with the findings by Bode et al.~\cite{bode2014human}. The lit exit sign treatment was always with the left exit except when both the sign and crowd treatments were conflicting (Crowd left and Sign right). Additionally, the participants made their decisions sooner, on average, when they chose to follow the lit exit sign treatment. Furthermore it was observed that as exit signs are expected to be lit, the unlit or dissuasive signs discouraged exit utilization and the lit signs encouraged exit utilization in general. The findings from these results agree with existing literature~\cite{olander2017dissuasive, galea2014experimental, galea2017evaluating}. 

The effectiveness of the lit signs to attract participants towards that particular exit and a similar effectiveness of the sign treatment when reinforced by the crowd treatment may suggest an increased tolerance for queuing and thus a cause for congestion when individuals trusted the information provided by the environment. 

\section{Limitations}
While the data is correct and useful for furthering the understanding of exit choice during an evacuation, it is important to recognize that there are limitations present from the recruitment pool and from the experiment set up itself. The participants were primarily recruited from the graduate student population of the University's College of Engineering which will likely not encompass the full range of the population demographics. Lastly, additional scenarios for the uneven crowd would have been beneficial for better understanding of crowd phenomena, such as the effects of number of people in the crowd. These can be addressed in future studies. Also, there will always be a question about the transferability of VR-based or screen-based study to real world evacuation situation. It is to be noted that VR provides a good compromise in terms of immersive experience and also a safe environment for data collection. 

\section{Conclusions}
The effect of lit exit signs and crowd movement on the exit choice behaviors of participants in an immersive VR experiment were studied. Specifically, the effect of an updated exit sign, different proportions of crowd movement toward each exit, and the effect of reinforcing and conflicting effects between the sign and crowd treatments were studied. Crowd and lit (updated) sign treatments were found to produce significant effects for following the treatment and time to initiate egress towards an exit. The sign treatment was found to be effective at reducing the decision time and increasing utilization of the exit with the lit sign. The utilization of the exit with the lit sign was not significantly reduced when the crowd treatment was reinforcing the sign treatment. Based on the results, the sign was an effective treatment and the crowd had an insignificant repulsive effect but was overall ineffective. These results agree with the literature. Specifically that the crowd can be repulsive~\cite{bode2013human,haghani2017social,haghani2019herding,lovreglio2014role} and a dissuasive sign can produces the intended effect~\cite{olander2017dissuasive, galea2014experimental, galea2017evaluating}. Furthermore, a potential learning effect was found, possibly due to the effect of the physicality of crowding more realistically perceived in immersive VR experiments. This may have also shown a difference between immersive VR and non-immersive screen-based experiments~\cite{bode2013human,bode2014human,bode2015information} because the participants are able to perceive the crowd more realistic in an immersive VR experiment they reacted to the crowd more strongly one way or another. 

Summarizing the implications of this work for human factors/evacuation planning/building design there is an indication that a queuing tolerance exists when crowds all move towards a trusted source (lit sign). Crowds are ineffective at encouraging utilization of a particular exit and may have a slight repulsive effect if the evacuee becomes uncomfortable with the crowding or feels delayed by the queuing.  A sign being updated through lighting/unlighting is an effective solution to influence exit choices. Particularly, when exit sign and crowd reinforced each other, the increased tolerance for queuing may causes increased congestion, thus a dynamic lighting of signage could be used to alleviate the crowding problem at bottlenecks/exits. This idea has potential to help handle a mass evacuation better in terms of average egress time for the crowd. In other words, dynamic signage has potential to help prevent over-crowding of exits which could in turn help with better flow of the crowd during emergency evacuation. This is an interesting direction for future research. 
\section*{Acknowledgment}
The authors would like to thank Mr. Hema Sumanth for his help with the data collection. This research was funded by NSF CPS Program (Award Number 1932505).
\bibliographystyle{elsarticle-num}

\bibliography{EDM}

\section*{Appendix}
\appendix
\section{Building Evacuation Experimenter Script}\label{sec:Appx1}

\begin{itemize}

\item Have the participant sign the informed consent form.

\textit{"Please read and sign the informed consent form. If you have any questions or concerns, then please let me know."}

\item Have the participant seat in seating area for the simulation.

\textit{"Please take a seat here."}

\item Instruct the participant of the controls and demonstrate with the controller. 

\textit{"The left joystick controls the movement and the right joystick controls turning."}

\item Explain the importance of the experiment. 

\textit{"We are testing evacuation behaviors during emergency situations to better provide safe evacuation outcomes by understanding what factors influence their evacuation decisions."}

\item Help the participant place the VR headset on their head.

\item Start simulation and perform VR preparatory Scenarios.

\item Explain VR preparatory experimental setup. 

\textit{"You will need to walk through the initial doorway and make a left or right decision, then walk through another doorway and make a left or right decision, and finally proceed until a final exit is reached."}

\item For the second set of the preparatory scenarios, explain to the participant that there will now be others who are also trying to exit.

\textit{"As far as you are aware, this scenario will be the same as the last except now others will also be trying to evacuate."}

\item After completing TR 2, have the participant take off the headset and complete survey 1 and the demographic form.

\textit{"Please remove the headset and complete this survey and this demographic form."}

\item Ensure that the participant has time to be outside of the headset and ask if they are okay to continue.

\textit{"Are you okay to continue?"}

\item If they are okay to continue, then inform them that there will be two exits from a room with provided exit information that they can choose to use. Also, upon their exit, they will receive a message stating “Success” for a successful exit or “Hazard” for an unsuccessful exit. Tell them to inform you if they have seen the message.

\textit{"There will be two exits from a room with provided exit information that they can choose to use."} 


\item After completing the scenarios, please have the participant take off the headset and perform survey 2.

\textit{"Please take off the headset and complete this survey."}

\item Thank the participant for contributing towards evacuation safety.

\textit{"Thank you for contributing towards evacuation safety."}

\end{itemize}

\end{document}